\documentclass[sigplan,10pt]{acmart}
\renewcommand\footnotetextcopyrightpermission[1]{}

\usepackage{graphicx}
\usepackage{balance,url,textcomp,amsmath}
\usepackage{tabularx,ragged2e,booktabs,caption}
\usepackage[linesnumbered,ruled]{algorithm2e}

\AtBeginDocument{%
  \providecommand\BibTeX{{%
    \normalfont B\kern-0.5em{\scshape i\kern-0.25em b}\kern-0.8em\TeX}}}
    
\def\Section {\S}

\newcommand{\squishlist}{
 \begin{list}{$\bullet$}
  { \setlength{\itemsep}{0pt}
     \setlength{\parsep}{3pt}
     \setlength{\topsep}{3pt}
     \setlength{\partopsep}{0pt}
     \setlength{\leftmargin}{1.5em}
     \setlength{\labelwidth}{1em}
     \setlength{\labelsep}{0.5em} } }

\newcommand{\squishlisttwo}{
 \begin{list}{$\bullet$}
  { \setlength{\itemsep}{0pt}
     \setlength{\parsep}{0pt}
    \setlength{\topsep}{0pt}
    \setlength{\partopsep}{0pt}
    \setlength{\leftmargin}{2em}
    \setlength{\labelwidth}{1.5em}
    \setlength{\labelsep}{0.5em} } }

\newcommand{\squishend}{
  \end{list}}
  
\renewcommand\footnotetextcopyrightpermission[1]{} 
\begin{document}

\title{Enabling Sustainable Clouds: The Case for Virtualizing the Energy System}
\author{Noman Bashir,$^1$ Tian Guo,$^2$ Mohammad Hajiesmaili,$^1$ David Irwin,$^1$ Prashant Shenoy,$^1$\\Ramesh Sitaraman,$^1$ Abel Souza,$^1$ and Adam Wierman$^3$
\\
\emph{$^1$UMass Amherst \hspace{3cm} $^2$WPI \hspace{3cm} $^3$Caltech}}
\date{}

\begin{abstract}
Cloud platforms' growing energy demand and carbon emissions are raising concern about their environmental sustainability.  The current approach to enabling sustainable clouds focuses on improving energy-efficiency and purchasing carbon offsets.  These approaches have limits:  many cloud data centers already operate near peak efficiency, and carbon offsets cannot scale to near zero carbon where there is little carbon left to offset.  Instead, enabling sustainable clouds will require applications to adapt to when and where unreliable low-carbon energy is available.   Applications cannot do this today because their energy use and carbon emissions are not visible to them, as the energy system provides the rigid abstraction of a continuous, reliable energy supply.  This vision paper instead advocates for a ``carbon first'' approach to cloud design that elevates \emph{carbon-efficiency} to a first-class metric. To do so, we argue that cloud platforms should \emph{virtualize the energy system} by exposing visibility into, and software-defined control of, it to applications, enabling them to define their own abstractions for managing energy and carbon emissions based on their own requirements. 

\end{abstract}

\maketitle
\pagestyle{plain}

\section{Introduction}
\label{sec:introduction}
Cloud platforms offer many advantages over building out and managing a private infrastructure, including low upfront costs, pay-as-you-go pricing, and rapid scalability.\footnote{Author list in alphabetical order by last name.} These advantages, along with an explosion of data and data-driven workloads~\cite{talia2013clouds,stevens2020ai,spark}, has fueled exponential growth in cloud capacity, which has been  doubling roughly every four years for more than a decade~\cite{srgresearch20}.  This growth has been accelerating recently due to increasing demand for AI and machine learning (ML) applications.  For example, recent estimates suggest the computation required to train state-of-the-art AI/ML models, e.g., for facial recognition, has been doubling every $3.4$ months for a decade, which is significantly faster than Moore's Law~\cite{openai}.  
If these trends continue, we can expect the cloud's exponential growth to persist for the foreseeable future, and possibly even accelerate further.


Cloud platforms have long had a strong financial incentive to optimize energy-efficiency to lower their operating expenses, which are massive for hyper-scale cloud data centers. These optimizations have been quite successful: despite the end of Dennard scaling~\cite{dennard}, the cloud's energy demand grew much more slowly than expected over the past decade~\cite{lbnl}.  The success was largely due to industry's intense focus on reducing data centers' power usage effectiveness (PUE) to near $1$ by aggressively optimizing energy-efficiency \emph{en masse} across hardware, software, and cooling systems.   However, there are few significant remaining opportunities left to further optimize energy-efficiency, as many cloud data centers already operate near peak efficiency~\cite{nsf-report}.  As a result, cloud platforms can no longer rely on improving their energy-efficiency to mitigate their energy growth. 
\emph{Thus, moving forward, the cloud's continued exponential growth is likely to translate directly into exponentially rising energy demand.}


As cloud platforms' energy demand grows, there is increasing concern about its environmental sustainability.   Indeed, researchers are gradually shifting their focus from reducing operating costs by minimizing energy usage to reducing environmental impact by minimizing carbon emissions.  For example, there has been recent work highlighting deep learning's carbon emissions~\cite{mccallum,stochastic-parrots} and advocating for the design of zero-carbon clouds~\cite{zero-carbon}.  Many prominent technology companies have also recognized the problem and set ambitious goals to reduce their carbon footprint, primarily by purchasing carbon offsets~\cite{amazon-carbon-neutral,facebook-carbon-neutral,vmware-carbon,google-carbon-free,microsoft-carbon-negative}.  However, carbon offsets are only a transitional mechanism that cannot scale to near zero carbon, since, at that point, there is little carbon left to offset.   Of course, the pronouncements above also coincide with a burgeoning financial incentive to ``go green,'' as solar energy is already the cheapest form of electricity in recorded history, and its cost is expected to continue to decline~\cite{cheap-article}.



This recent emphasis on sustainability  recognizes that the cloud's rising energy consumption is not actually the problem: rather, \emph{the problem is the carbon footprint of that energy consumption and its negative impact on the environment.}  However, optimizing data centers for \emph{carbon-efficiency}, i.e., the work done per kilogram of carbon emitted, differs substantially from optimizing them for \emph{energy-efficiency}, i.e., the work done per joule of energy consumed.  To illustrate, a data center could be highly energy-inefficient, e.g., PUE$\gg$$2$, but also highly carbon-efficient if it is powered entirely by \emph{green} energy from co-located renewables with zero carbon emissions. Likewise, a data center could be highly energy-efficient, e.g., PUE$\sim$$1$, but also highly carbon-inefficient if it is powered by \emph{brown} energy from grid generators burning fossil fuels.  While there has been decades of research on optimizing cloud energy-efficiency, there has been little research on optimizing cloud carbon-efficiency.  To address the problem, we advocate for a ``carbon first'' approach to cloud design that re-focuses research on optimizing carbon-efficiency by elevating it to a first-class metric.

\begin{figure}[t]
    \centering
    \includegraphics[width=0.5\textwidth]{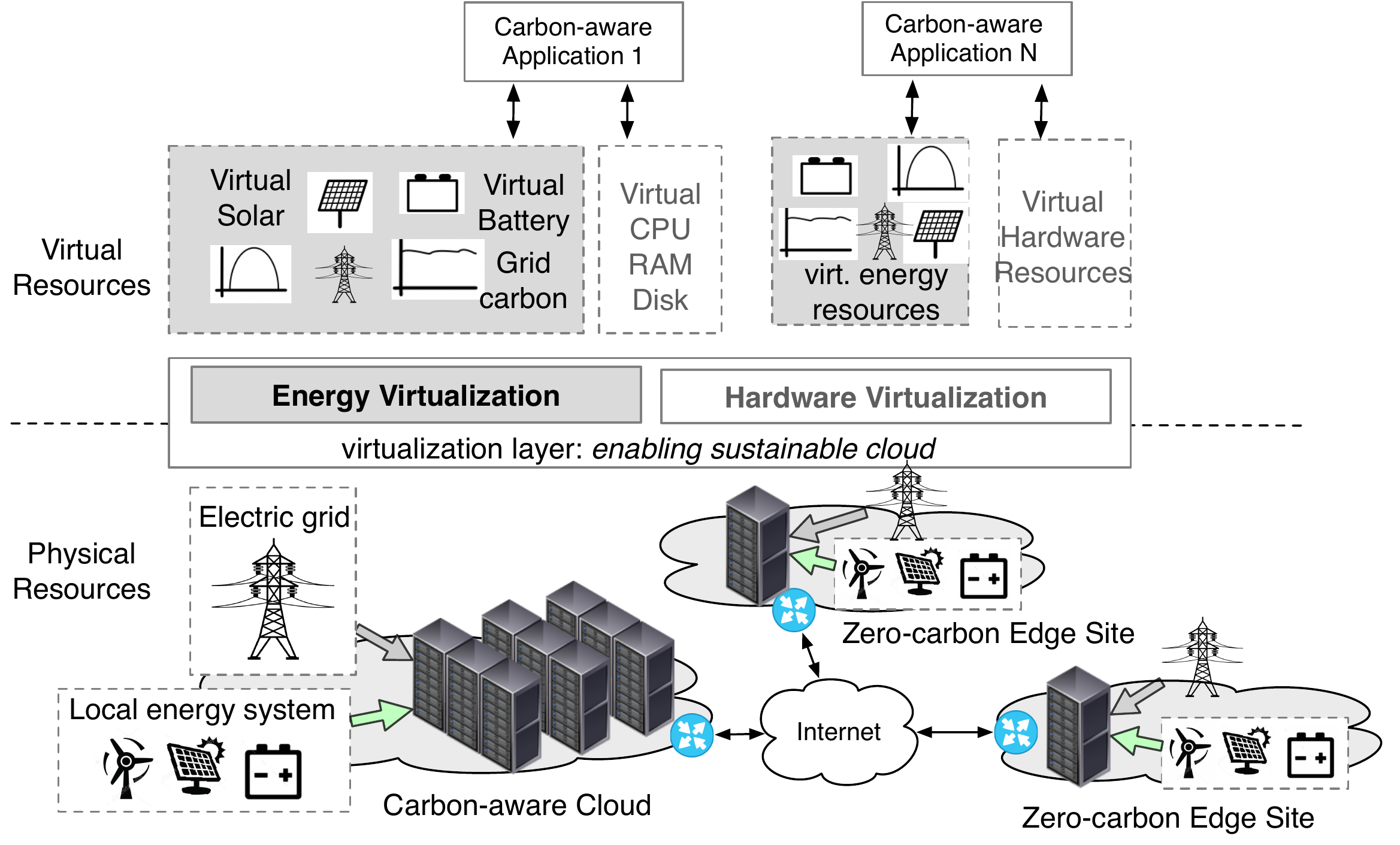}
    \vspace{-0.3cm}
    \caption{\emph{Our vision of a sustainable cloud-edge infrastructure that virtualizes the energy system, enabling applications to optimize their own energy usage and carbon emissions.}}
    \vspace{-0.4cm}
    \label{fig:overview}
\end{figure}

A distinguishing characteristic of low-carbon energy is that it is \emph{unreliable}: it is not always available at any single location all the time, but instead varies widely, and not entirely predictably, both temporally and geographically. 
Thus, \emph{enabling sustainable clouds that are carbon-efficient will require applications to adapt to when and where low-carbon energy is available.} Unfortunately, applications cannot do this today because their energy and carbon are not visible to them, as the energy system provides the rigid abstraction of a continuous, reliable energy supply.  To address the problem, this vision paper instead
argues for \emph{virtualizing the energy system} by exposing visibility into, and software-defined control of, it to cloud applications, providing them the flexibility to define their own abstractions for managing energy and carbon emissions based on their own requirements.  

Figure~\ref{fig:overview} illustrates our vision of a sustainable cloud-edge infrastructure that virtualizes the energy system, enabling applications to optimize their energy usage and carbon emissions. 
Our approach effectively extends the end-to-end principle to the energy system~\cite{endtoend}, and to some extent takes prior approaches that advocate delegating resource management to applications to their logical conclusion~\cite{exokernel,mesos}.
\section{Current Approaches Are Not Sustainable}
\label{sec:background}
Ironically, current approaches to enabling sustainable clouds, which focus on improving energy-efficiency and purchasing carbon offsets, are themselves not sustainable.  That is, as we discuss below, these approaches are not sufficient to eliminate the cloud's carbon emissions, and it is not necessarily clear how much they contribute to lowering them, especially over long periods.  Thus, developing new approaches that unequivocally and directly reduce carbon emissions is critical. 



\noindent {\bf Energy-Efficiency}.  Improving the energy-efficiency of computer systems, in general, has been an active research area for nearly three decades~\cite{weiser-dvfs}, and of data centers, in particular, for at least two decades~\cite{muse}.  Over this period, researchers have developed myriad techniques that have pushed computational energy-efficiency to near its physical limits.  These techniques run the gamut from hardware mechanisms, such as dynamic voltage and frequency scaling (DVFS)~\cite{weiser-dvfs} and power capping~\cite{power-capping,rapl}, to software policies, such as consolidating workload to shutdown idle servers~\cite{muse}, to facility and cooling system optimizations, such as leveraging ``free'' cooling~\cite{free-cooling1} and optimizing power delivery~\cite{power-routing}.  Improvements in the latter are captured by PUE, and have been an intense focus of data center operators over the past decade, largely because PUEs were initially high ($\gg$2) and thus there was significant room for improvement. 


At this point, however, state-of-the-art data centers have PUEs within 6-10\% of optimal, i.e., PUE$=$$1$~\cite{google-pue}, leaving little room for further improvement.  Hardware and software systems similarly operate near their peak efficiency and thus also cannot be substantially improved.  As a result, \emph{we cannot significantly reduce the cloud's energy consumption and thus carbon emissions by further improving energy-efficiency.}   In fact, the efficiency improvements above have not even stopped the cloud's energy growth, but only slowed it down relative to initial estimates~\cite{koomey1}.  This may be due, in part, to Jevon's Paradox, which observes that increasing resource efficiency often increases, rather than decreases, that resource's consumption~\cite{jevons1,jevons2}.   This paradox is well-known in energy economics, and occurs because increasing energy's efficiency also lowers its cost, which can increase demand, i.e., usage, such that it more than offsets the decrease in usage from improving efficiency.  Hence, while there are many benefits to improving energy-efficiency, reducing energy consumption, and thus carbon emissions, is not always one of them. 

\noindent {\bf Carbon Offsets}. The other common approach to reducing cloud carbon emissions  is by purchasing carbon offsets, which represent a quantifiable reduction in carbon emissions. While there are many types of carbon offsets, technology companies often purchase them by either subsidizing renewable energy projects or purchasing renewable energy credits~\cite{google-ppa}.  This renewable energy is not necessarily located near cloud data centers, and typically does not directly power them.  However, when accounting for carbon emissions, these carbon offsets are treated as location-agnostic by assuming each kilowatt-hour (kWh) of carbon-free renewable energy displaces a kWh of carbon-intense electricity consumed from data centers' local grid. The typical accounting period for carbon offsets under GHG Prototcol is a calendar year~\cite{ghg}.  Thus, a ``carbon neutral'' company purchases enough carbon offsets in a year to balance its carbon emissions.   Using this approach, companies can also be ``carbon negative'' by purchasing more carbon offsets than their carbon emissions~\cite{microsoft-carbon-negative}.


Technology companies have led in the adoption of carbon offsets, and many have used them to achieve annualized location-agnostic carbon-neutrality~\cite{amazon-carbon-neutral,facebook-carbon-neutral,vmware-carbon}.  Yet, these and other companies' operations are still responsible for a significant amount of direct carbon emissions. To address the problem, Google recently announced that it aims to be ``carbon free'' by 2030, in part, by piloting a new form of carbon offset, called Time-based Energy Attribute Certificates (T-EACs), which have an hourly location-specific accounting regime~\cite{google-teac}.  T-EACs recognize that grid carbon emissions vary over time and by location, based on the mix of generators used to satisfy demand, and thus incentivize using energy when and where carbon emissions are low.  As a result, T-EACs more directly offset companies' direct carbon emissions.  Unfortunately, carbon offsets, including T-EACS, are only effective as a transitional mechanism, since we must ultimately reduce \emph{absolute} global carbon emissions to near zero, where there is little carbon left to offset.   \emph{To reach zero carbon, we must eventually move beyond carbon offsets and focus on changing operations to always run on zero-carbon energy, e.g., from solar, wind, hydro, nuclear, geothermal, etc.}   Carbon offsets disincentive, and may actually delay, these operational changes by providing a means for reducing carbon emissions for less than it would cost to make such changes~\cite{wired}.

\section{A Sustainable Carbon First Approach}
\label{sec:overview}


To address the limitations of current approaches, we advocate for a ``carbon first'' approach that elevates \emph{carbon-efficiency} to a first-class metric in cloud design. Carbon-efficiency is a measure of computational work done per kilogram of carbon, and other greenhouse gas (GHG), emissions.  We focus specifically on Scope 2 emissions from using electricity~\cite{ghg}, which represent the vast majority of cloud platforms' carbon emissions.  Unlike energy-efficiency, carbon-efficiency optimizations are not bound by Jevon's Paradox because carbon is a not a resource cloud platforms consume, but, rather, an energy by-product, which they can eliminate by operating when and where zero-carbon energy is available. 


Of course, to optimize any metric, we must be able to both measure and control it.  Currently, cloud platforms and users have little visibility into their energy usage and carbon emissions, and little-to-no control over them.  The cloud's energy system delivers power from two primary sources:  the electric grid and an increasingly rich local energy system, which may consist of on-site renewables, such as solar, and batteries. 
These two power sources present different trade-offs. The grid provides the convenient abstraction of a reliable power source with a non-zero carbon footprint, while the local energy system is an unreliable source of  clean  power.   
In \Section\ref{sec:grid} and \Section\ref{sec:local}, we discuss the potential for optimizing carbon-efficiency by exposing visibility and control of energy and carbon emissions from both sources.   \Section\ref{sec:virtualize} then makes the case for exposing this visibility and control directly to cloud applications by virtualizing the energy system. 

 
\subsection{Electric Grid: Visibility and Control}
\label{sec:grid}


The electric grid presents electrically-powered devices with the abstraction of a reliable supply of power on demand (up to some maximum power).   The grid exposes a hardware interface to this abstraction in the form of standardized electrical sockets, which implicitly encode information about power's output, i.e., voltage, frequency, etc., via their form factor.  This simple abstraction has been amazingly successful for over 100 years, since the grid's inception, at supporting countless uses of electrical energy, most of which were unimaginable a century ago.  However, this simplicity is increasingly becoming a barrier to innovation, as it exposes no interface for receiving information about grid energy's characteristics, i.e., its energy sources and carbon emissions, and no means for controlling these characteristics.  This lack of visibility and control prevents systems from optimizing carbon-efficiency. 


\noindent {\bf Exposing Visibility}.   The grid's energy comes from a large mix of generators that have a wide range of carbon emissions. For example, hydroelectric power plants have zero carbon emissions, while carbon emissions from thermal generators vary widely based on their fuel type, capacity, and real-time power output.  The grid varies its set of active generators, and their real-time power output, over time to match the energy demand of its connected devices.   Since the grid exercises little-to-no control over the devices that connect to its socket interface, it also has little-to-no control over variations in its energy demand.  \emph{These variations in demand and generation cause wide variations in grid energy's carbon emissions.}


Recently, new services, such as electricityMap~\cite{electricity-map}, have emerged that estimate grid carbon emissions by collecting and analyzing real-time data on grid operations published by grid balancing authorities.  
This data reveals that generator operations exhibit an order of magnitude variation in carbon emissions across space and time. Such spatial and temporal variations present an untapped opportunity for reducing the carbon emissions of cloud platforms and applications. 

\begin{figure}
\centering
\begin{tabular}{c}  
\includegraphics[width=0.42\textwidth]{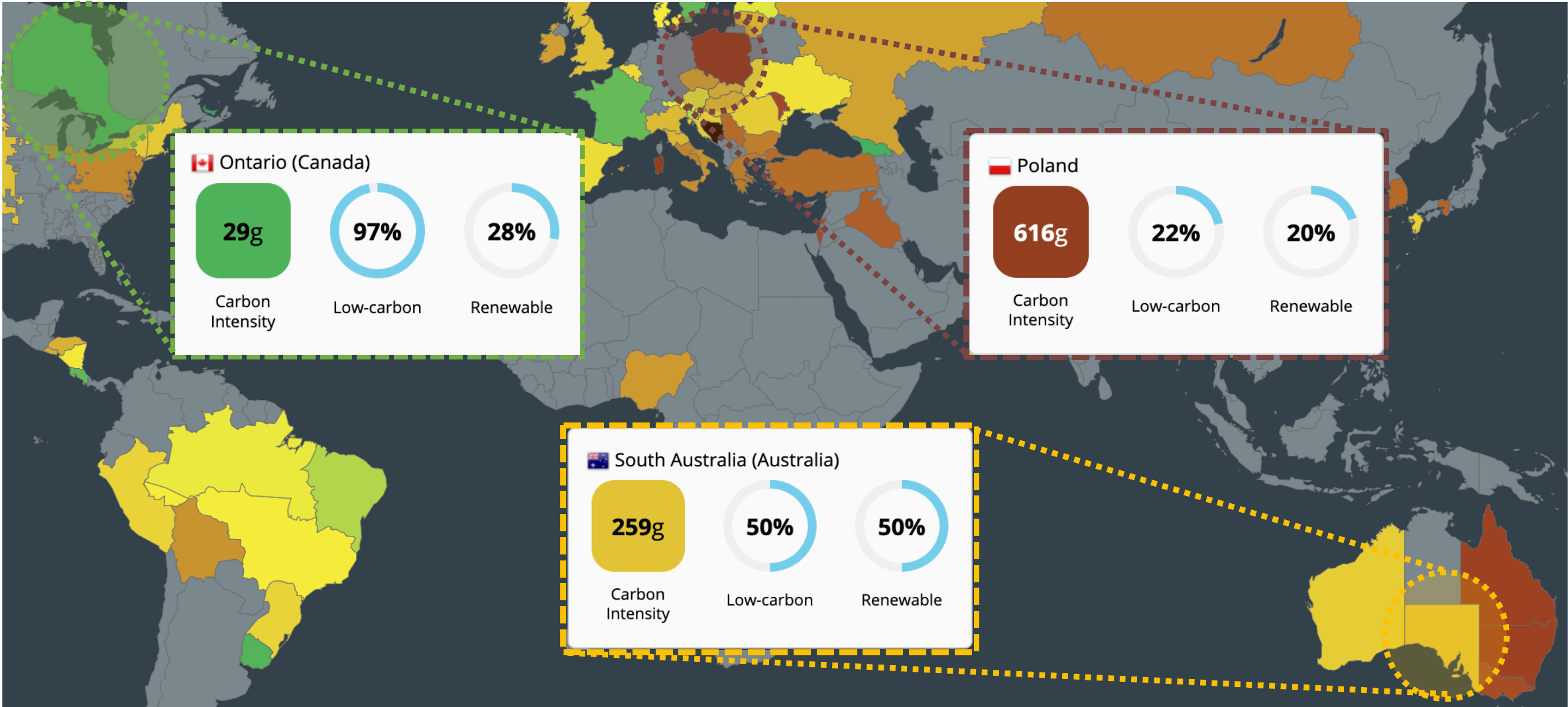}\\
(a) Grid carbon emissions vary by 21$\times$ across regions.\\
\includegraphics[width=0.42\textwidth]{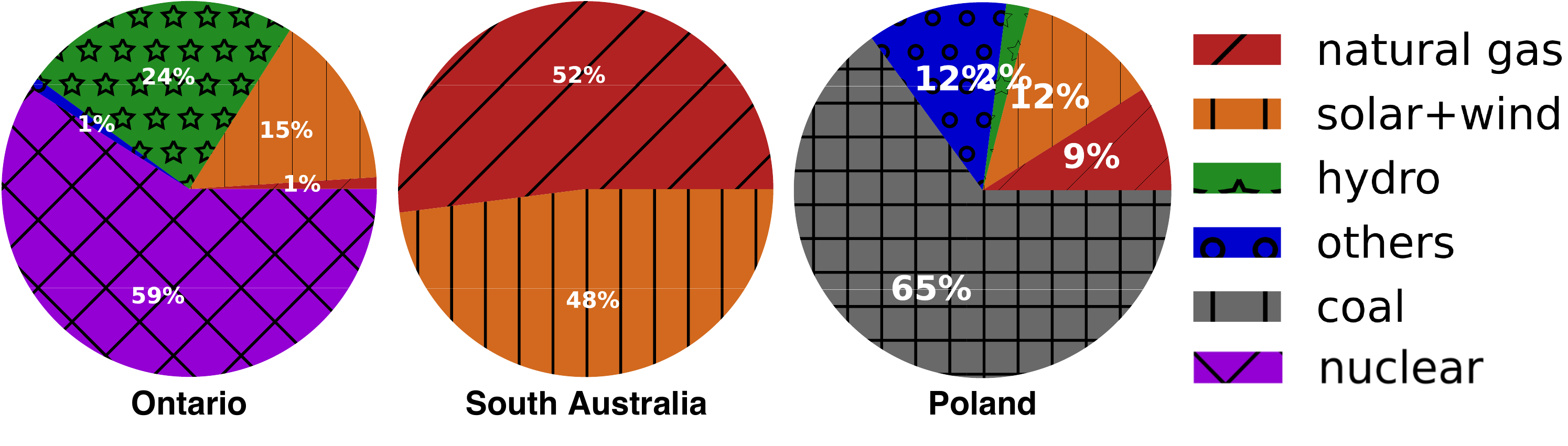}\\
(b) Varying mix of generators in different regions\\
\includegraphics[width=0.43\textwidth]{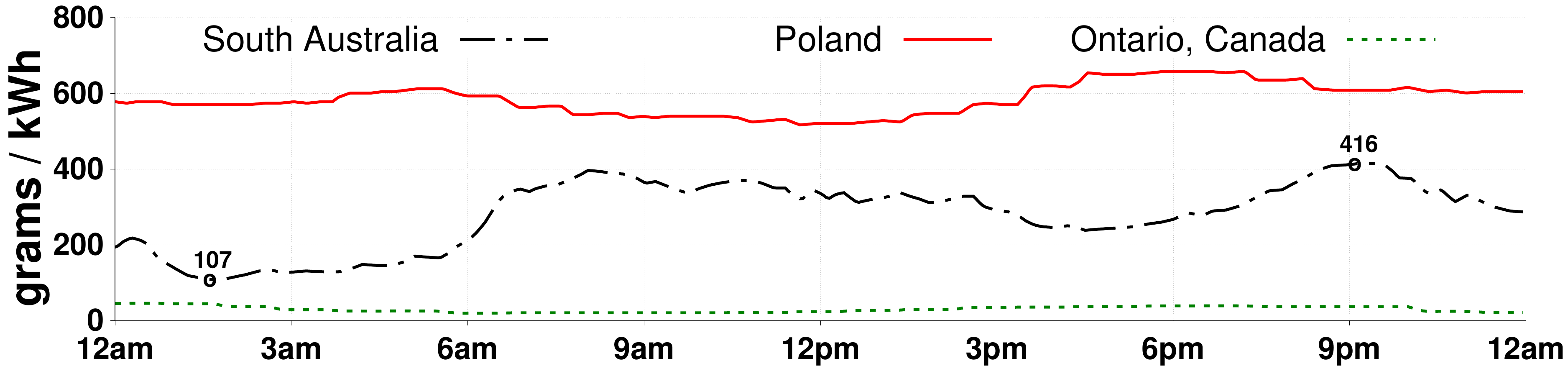} \\
(c) Hourly carbon emissions vary by 4$\times$ over a single day.
\end{tabular}
\vspace{-0.4cm}
\caption{\emph{Grid carbon emissions vary geographically (a) and temporally (c) based on the mix of active generators (b).}
}
\label{fig:carbon}
\vspace{-0.8cm}
\end{figure}

\emph{Variations Across Space.} Figure~\ref{fig:carbon}(a) shows the average carbon intensity (in g$\cdot$CO2/kWh) of different world regions from electricityMap, where darker colors indicate higher carbon intensity.   The variations are significant with Ontario, which relies on nuclear and hydro power, having 21$\times$ lower carbon intensity than Poland, which relies on thermal generators that burn coal.  Figure~\ref{fig:carbon}(b) shows the different generator mixes in Ontario,  S. Australia, and Poland from left to right.



\emph{Variations Across Time.} Figure~\ref{fig:carbon}(c) similarly shows that grid energy's carbon intensity also varies significantly over time at any single location.  On this day, S. Australia's carbon intensity varied by 4$\times$ (between 2am and 9pm). 

%


In some sense, electricityMap and similar services supplement the grid's socket interface with a new interface that enables devices to retrieve energy information. Thus, \emph{we can use these services to also expose grid carbon emissions to cloud platforms and applications.}   ElectricityMap already exposes a real-time API that software could integrate programmatically.  Google has recently taken steps in this direction by using electricityMap data to provide users carbon intensity estimates for cloud regions to inform their choices~\cite{google-carbon}. Interestingly, Google and other cloud platforms do not yet expose similar visibility into cloud applications' fine-grained power usage even though such information is typically visible to low-level software.  System software, such as RAPL~\cite{rapl}, is also capable of attributing power to individual processes, containers, and virtual machines (VMs) on a server. Thus, while Google's current carbon data is useful for coarse-grained region selection, users cannot yet track their own fine-grained energy usage, and thus carbon emissions.  
\emph{We argue that exposing such fine-grained data is necessary for  optimizing the carbon emissions of  cloud platforms and  applications.}

\noindent {\bf Exposing Control}.  
Exposing the visibility above would enable direct monitoring of grid energy's carbon emissions, but is useless without a means to control these emissions.   Such control is possible either by changing energy's supply, i.e., the grid's mix of generators, or its demand, i.e., consumption.  Unfortunately, consumers cannot directly influence grid energy's carbon emissions by altering the mix of generators supplying electricity.  \emph{Consumers, including cloud platforms, can, however, vary their energy consumption, and hence carbon emissions, by shifting workloads in either time---to a low-carbon period---or space---to a low-carbon region.}   

Cloud platforms can shift energy usage across time by controlling either computation via job scheduling, i.e., running jobs during low-carbon periods, or energy via battery scheduling, i.e., charging and discharging batteries during low- and high-carbon periods, respectively. The latter is important for interactive services that cannot simply shift their computation in time.  Google has begun experimenting internally with such carbon-aware time-shifting in its data centers~\cite{google-carbon2}.  Cloud platforms can similarly shift energy usage across space by moving computation, e.g., jobs, requests, etc., to regions with low carbon intensity.    As others have noted~\cite{terrawatt}, moving computation, which is equivalent to moving tiny amounts of energy in the form of the bits that encode it, is many orders of magnitude more efficient and less costly than moving energy.  Thus, \emph{cloud platforms should focus on moving computation to energy, rather than moving energy to computation.} Cloud platforms' geographically distributed infrastructure enables such carbon-aware movement.  

Unfortunately, cloud platforms do not expose any means for controlling energy usage (and thus carbon emissions), such as power capping~\cite{power-containers}, even though these mechanisms are also provided by low-level software, including RAPL~\cite{rapl}, and can be applied to individual processes, containers, and VMs.  There has been substantial prior work using such control to perform time and space shifting but mostly for optimizing energy costs, which, unlike carbon emissions, are visible, e.g., via job scheduling~\cite{parasol,greenslot}, battery scheduling~\cite{dr2,bhuvan2}, and request routing~\cite{cutting}.  Some of these approaches may also be useful in optimizing cloud carbon-efficiency.  However, a key difference with prior work is that we eventually must reduce carbon emissions to zero. This will likely require capping carbon emissions at some point, and then progressively lowering the cap.   Such caps will break the grid's abstraction of reliable energy and force cloud platforms and applications to handle it. In contrast, prior work generally focuses on minimizing cost assuming reliable energy, and thus does not address problems with unreliable energy. As we discuss in \S \ref{sec:local}, techniques designed for transient computing~\cite{yank,flint}, which focuses on designing reliable applications on unreliable servers, may be useful in handling this unreliability. 




\subsection{Local Energy System: Visibility and Control}
\label{sec:local}

Cloud platforms can also draw energy from their local energy system, which may include both substantial battery capacity and co-located renewables, such as solar or wind energy~\cite{maiden}.    Unfortunately, hyper-scale cloud data centers are generally too power-dense for renewables to power a significant fraction of their operations.  For example, the average solar power density across the earth's surface is $\sim$$+55$W/m$^2$~\cite{solar-stats}, assuming ideal solar efficiency at the Shockley-Queisser limit. Thus, assuming average solar density, making a 100MW data center net-zero using co-located solar would require covering an area of nearly 450 acres (or $\sim$$1.35$km$\times$$1.35$km area).  
While such a large co-located solar farm may be possible~\cite{maiden}, there are many issues that can prevent it, e.g., existing land ownership, unsuitable terrain, distance to grid connection, etc.  As a result, we expect that most hyper-scale data centers will require some grid energy, and thus will need to rely, in part, on their local grid to reduce carbon intensity.






The characteristics above, though, motivate more widely distributing computation to lower its power density to match renewables, rather than aggregating it at a single location.   Hence, we envision sustainable clouds will include many smaller, more widely distributed, micro data centers that are largely self-powered, primarily using solar and batteries~\cite{parasol}.  This approach, while more sustainable, also has some financial benefits.  As mentioned earlier, moving computation is cheaper than moving energy, and the difference in cost will only grow as solar prices continue to decline.  The cost disparity is not apparent now, in part, because the grid already exists, and thus represents a sunk cost.  Cloud platforms have also already started building out such a distributed infrastructure, albeit for a different reason:  to support edge computing near end-users that can deliver low-latency services~\cite{stadia}. \emph{Cloud platforms should leverage this opportunity to also improve their sustainability.}  Prior work has already demonstrated that renewable-powered edge sites are viable~\cite{parasol}.

\begin{figure}[t]
    \centering
    \includegraphics[width=0.48\textwidth]{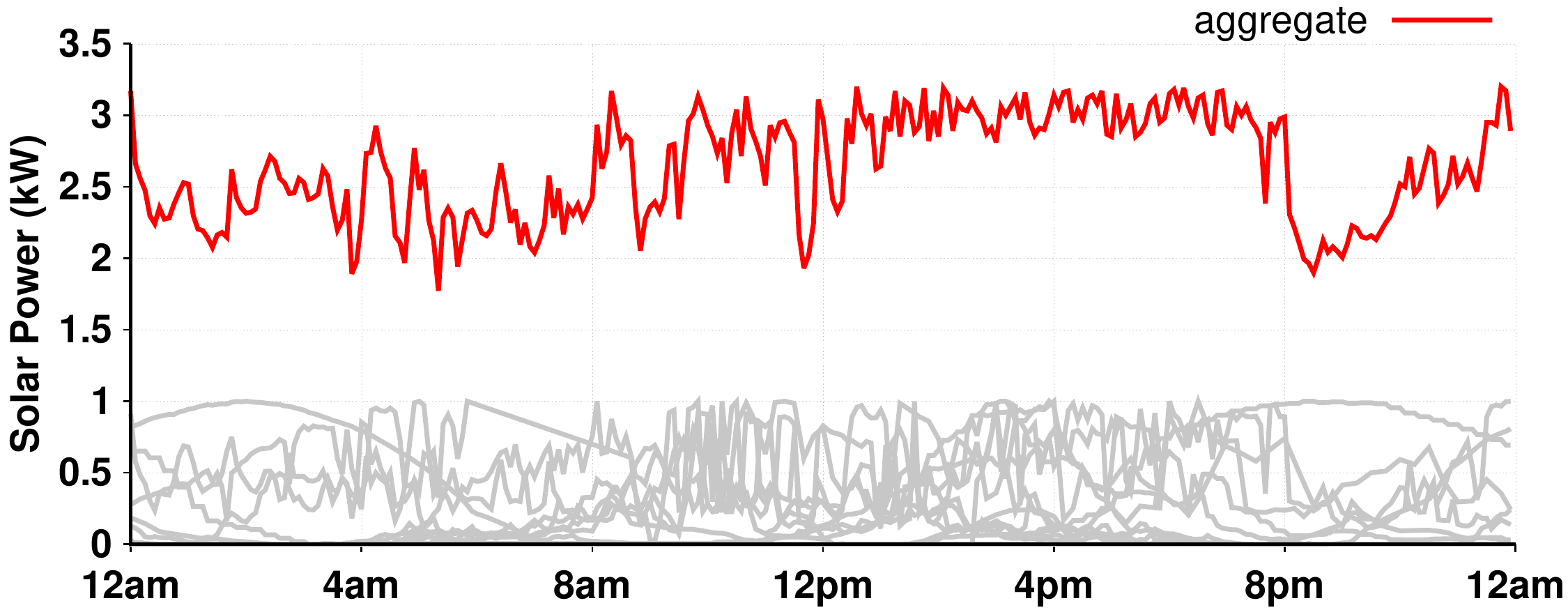}
    \vspace{-0.3cm}
    \caption{\emph{Estimated individual and aggregate solar power across 16 AWS regions distributed across the earth.}}
           \vspace{-0.4cm}
    \label{fig:aws-aggregate}
\end{figure}

\begin{figure}[t]
    \centering
    \includegraphics[width=0.48\textwidth]{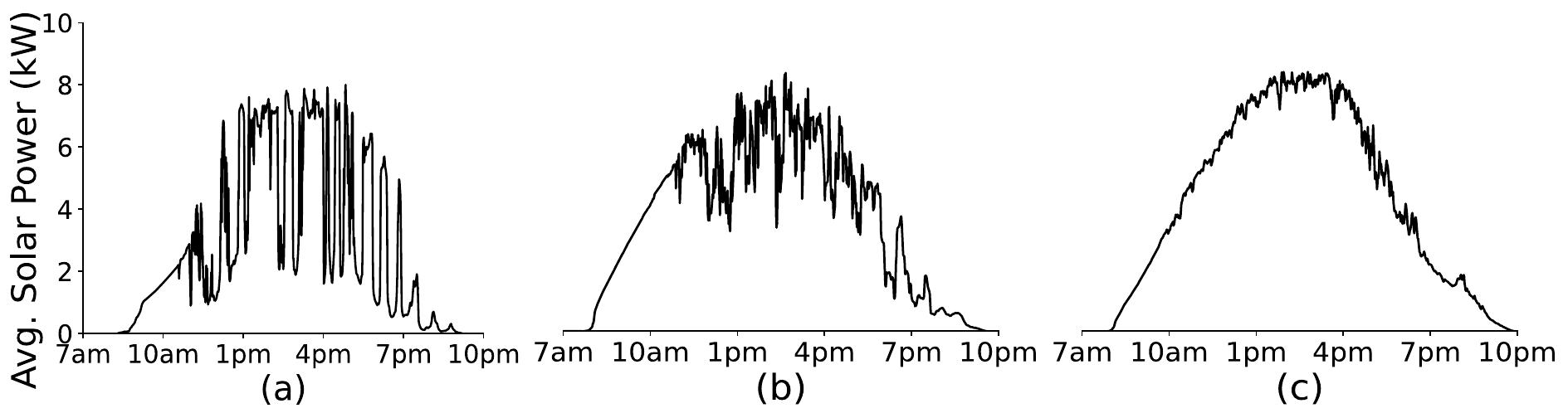}
    \vspace{-0.3cm}
    \caption{\emph{Solar power aggregated across (a) 1, (b) 7, and (c) 15 sites, from left to right, (8kW each) within a 50-mile radius.}}
        \vspace{-0.3cm}
    \label{fig:solar-aggregate}
\end{figure}


Distributing multiple such edge sites across large and small geographical regions is also beneficial because it can reduce renewable energy's unreliability.  Consider that, over large geographic regions, solar energy is highly reliable: in the extreme, the earth's surface receives near constant solar radiation.   This effect is also present when applied to just a few sites, as shown in Figure~\ref{fig:aws-aggregate}, which plots estimated solar output across 16 AWS regions spread across the world (gray), and their aggregate (red).  The figure shows that aggregate solar output is much less volatile than any individual site.  Solar volatility also decreases when aggregating across many sites in a small region, as shown in Figure~\ref{fig:solar-aggregate}.  This figure shows aggregate solar energy, from left to right, across 1, 7, and 15 sites (each with 8kW capacity) over a cloudy day within a 50-mile radius.  As before, the aggregate is much less volatile than a single site. Thus, \emph{while renewable energy is not always available at one location, it is likely available somewhere. }




\noindent {\bf Exposing Visibility}.  Self-powered edge data centers have complete visibility into their rich local energy system, including its real-time solar (or wind) generation, grid power consumption, battery charge level, and energy usage of computing infrastructure. These sub-systems generally include fine-grained energy monitoring that report power consumption at high resolutions, e.g., every second or minute, using embedded sensors or, for servers, hardware counters.  This energy-related information is typically available via programmatic APIs exposed by different hardware components, e.g., battery charge controller, solar inverter, etc., and should be exposed and integrated into management software to provide applications visibility into the energy system.

\noindent {\bf Exposing Control}.  Although local generation from renewable sources is intermittent and variable, edge data centers have full  control over their local energy system, enabling them to  dynamically coordinate power flow between computing infrastructure and the grid, solar, and batteries~\cite{parasol}.  Hyper-scale data centers with co-located renewables have similar control, but as mentioned above, are more likely to rely on grid energy.  Specifically, data centers can control: when to use, store, or net meter solar energy; when and how to charge batteries, e.g., from the grid or solar, and when to discharge them; and if and when to use grid energy and how much.  Renewable energy research  has shown that such control decisions can directly optimize carbon emissions~\cite{emissions}. 


However, a key challenge for renewable-powered infrastructure is its unreliability: if grid energy is not available, e.g., due to carbon capping, it may require throttling or shutting down servers due to a lack of  energy.  Relying more on unreliable renewable energy will further break the grid's reliability abstraction.  Instead, carbon-efficient applications will need to be designed to handle new renewable energy dynamics.  These dynamics are reminiscent of those exhibited by spot/preemptible cloud servers~\cite{aws-spot,google-preemptible,azure-spot}. While spot/preemptible servers are cheap, cloud platforms may revoke them at any time. These revocation ``failures'' are expected, and prior work has designed many techniques to mitigate their impact in various applications, primarily by judiciously checkpointing in-memory state~\cite{yank,spotcheck,spoton,flint,tr-spark,pado}. 

An important difference between spot/preemptible dynamics and renewable dynamics is that, with the former, cloud platforms unilaterally choose which servers to revoke without warning, while, with the latter, systems have a wider range of choices to satisfy energy drops by controlling either their computation, e.g., selecting servers to throttle/shutdown, or their energy system, e.g., discharging batteries.  Of course, under an energy shortage, applications might also choose to move computation to another site without a shortage.  Different applications will make different decisions based on their own specific requirements.

\section{Virtualizing the Energy System}
\label{sec:virtualize}




The previous section argues that optimizing carbon-efficiency requires being able to measure and control it.   There is really no technical barrier to doing so, as \Section\ref{sec:overview} outlines a rich set of existing mechanisms for exposing visibility and control of energy usage and carbon emissions to cloud platforms.  Unfortunately, simply exposing visibility and control is not enough, since cloud platforms are not in a position to exercise this control to optimize carbon-efficiency because they have little visibility into the applications that run on them.   Applications have a wide range of characteristics and performance requirements that affects how they may choose to optimize carbon-efficiency.  For example, a batch job, such as training a large distributed ML model, might choose to handle a renewable energy shortage (or high-carbon grid energy) by capping its power usage if it can tolerate some delay. In contrast, an interactive service might handle a similar situation by either discharging a battery, re-routing requests to a site with ample clean energy, or some combination. 

Thus, we argue that enabling sustainable clouds requires \emph{virtualizing the energy system} to expose the visibility and control from \Section\ref{sec:overview} directly to applications.  An application's virtual energy system might provide a virtual solar array, which supplies a configurable share of a physical solar array's real-time power, a virtual battery, which offers a configurable share of a physical battery's capacity, and a grid connection, which provides access to a configurable amount of carbon-intensive energy. An exogenous policy would determine each application's share of variable solar power and battery capacity at each site.  For example, public cloud platforms might sell solar and battery shares at each site for some price independently of hardware resources.  Cloud platforms could also directly incentivize carbon-efficiency by setting per-user carbon caps or placing an explicit price on carbon. 

\begin{figure}[t]
    \centering
    \includegraphics[width=0.48\textwidth]{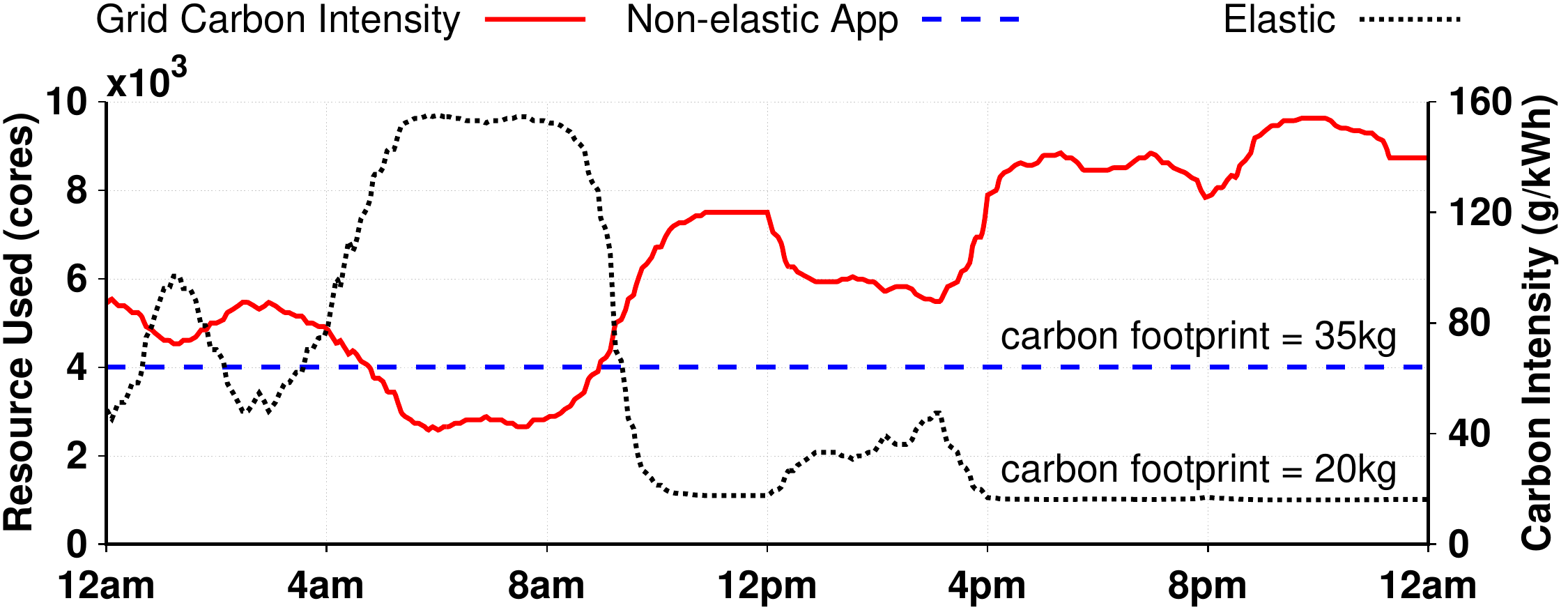}
    \vspace{-0.3cm}
    \caption{\emph{Resource usage and carbon intensity for an \emph{elastic} and \emph{inelastic} ML job.  The carbon-aware \emph{elastic} job finishes at the same time, and reduces carbon emissions by 45\%.}}
    \vspace{-0.5cm}
    \label{fig:carbon-adaptive-exp}
\end{figure}

\begin{figure}[t]
    \centering
    \includegraphics[width=0.48\textwidth]{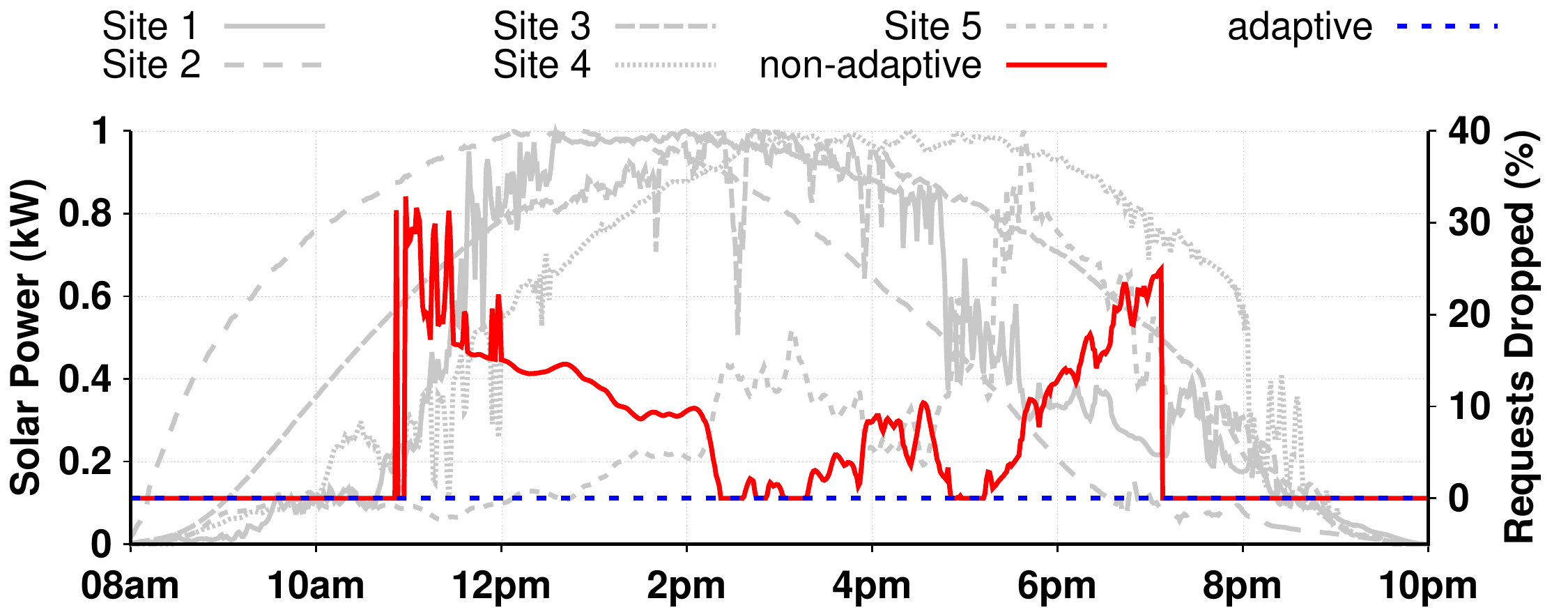}
    \vspace{-0.3cm}
    \caption{\emph{Solar power across multiple edge sites (gray lines), and requests dropped by adaptive and non-adaptive serverless edge application (blue and red lines).  The non-adaptive service drops up to 32\% of requests due to lack of renewable energy.}}
    \vspace{-0.3cm}
    \label{fig:energy-adaptive-exp}
\end{figure}

Critically, virtualizing the energy system would enable \emph{individual applications} to measure, control, and thus optimize their own carbon-efficiency.  Applications would have full control over their virtual energy system via an application-level interface, enabling them to explicitly regulate grid power usage, battery charging/discharging, e.g., from solar or grid, per-VM (or container) power caps, etc.  This interface would also enable applications to access energy information, such as their real-time energy usage, grid carbon emissions, and solar generation, or receive asynchronous notifications when 
significant changes occur.  Applications would then be responsible for matching their allocated energy's supply to demand to optimize their carbon-efficiency.  That is, applications would be allocated computing resources independently of energy, and would use their virtual energy system to explicitly allocate their available power across these resources, e.g., by setting power caps or deferring/moving computations.  The system would aggregate and map many applications' virtual controls onto the physical system, e.g., by enforcing power caps, regulating battery charging/discharging, etc.  Our approach essentially applies the end-to-end principle to the energy system~\cite{endtoend}, similar to how exokernels applied it to managing hardware in the OS~\cite{exokernel}. 

To demonstrate the benefits of such flexibility, Figure~\ref{fig:carbon-adaptive-exp} depicts the execution over time of elastic and inelastic distributed ML training, along with grid energy's variable carbon-intensity.   In this case, we use MLperf~\cite{mlperf} power-performance results to model ImageNet~\cite{deng2009imagenet}.  The inelastic variant uses the same resources, e.g., GPUs, throughout the day, while the elastic variant has an autoscaler that uses a sigmoid function to dynamically scale resources up and down during low and high carbon periods, respectively.  As a result, the elastic carbon-aware variant lowers carbon emissions 45\% without any effect on job completion time.


Figure~\ref{fig:energy-adaptive-exp} depicts the performance of a serverless edge application that runs on edge sites powered by renewable energy.  Requests are generated based on the latency-sensitive serving tasks in a Google workload trace (first day of cell a)~\cite{clusterdata:Wilkes2020a}.  The gray lines depict solar output at each site, while the blue and red lines depict the percentage of requests dropped by an adaptive and non-adaptive application.  For the non-adaptive application, we evenly and statically divide the workload across sites, while the adaptive applications routes workload to sites based on energy availability.  We report the drop rate only when the aggregate power across sites is enough to serve the load. The figure shows that the non-adaptive service drops up to 32\% of requests due to lack of energy, even though energy is available somewhere, while the adaptive service never drops requests and has a zero carbon footprint.

\section{Conclusions}
\label{sec:conclusions}

This paper advocates for a ``carbon first'' approach to cloud design that elevates carbon-efficiency to a first-class metric. Carbon-efficiency has generally been ignored by computer systems researchers in the past because it requires tighter integration with, and visibility into, an often opaque energy system that provides the convenient abstraction of reliable energy on demand. However, the environmental cost of maintaining this simple abstraction has become too high, as it masks energy's unreliability and carbon emissions from applications.  Thus, we argue for virtualizing the energy system to expose the visibility and control applications need to optimize carbon emissions based on their own requirements. 
\\
\noindent {\bf Acknowledgements.} This work is funded by NSF grant CNS-2105494 and VMware.

\balance
\bibliographystyle{ACM-Reference-Format}
\bibliography{paper.bib}

\end{document}